\documentclass[aps, prl, reprint, groupedaddress, amssymb, amsmath, amsfonts, superscriptaddress, 30pt]{revtex4-2}
\usepackage{graphicx} 
\usepackage{dcolumn}
\usepackage{xcolor}
\usepackage{soul}
\usepackage{ulem}
\usepackage{bm}
\usepackage[utf8]{inputenc}    
\usepackage[T1]{fontenc}   
\usepackage{capt-of}
\usepackage{float}
\usepackage{etoolbox}
\usepackage{siunitx}
\usepackage{booktabs}
\usepackage{multirow}
\usepackage{hyperref}
\usepackage[capitalise,nameinlink]{cleveref} 
\usepackage{etoolbox}
\usepackage{ragged2e} 
\usepackage{lineno}
\usepackage{CJKutf8}

\makeatletter
\AtBeginEnvironment{thebibliography}{%
  \let\href\@secondoftwo
}
\makeatother

\begin{document}

\begin{CJK*}{UTF8}{gbsn}

\title{Epigenetic state encodes locus-specific chromatin mechanics}

\author{Guang Shi (时光)}
\email{guang.shi.gs@gmail.com}
\affiliation{Department of Chemistry, The University of Texas at Austin, Austin, Texas 78712, USA}
\author{D. Thirumalai}
\email{dave.thirumalai@gmail.com}
\affiliation{Department of Chemistry, The University of Texas at Austin, Austin, Texas 78712, USA}
\affiliation{Department of Physics, The University of Texas at Austin, Austin, Texas 78712, USA}

\date{\today}

\begin{abstract}
Chromatin is repeatedly deformed \textit{in vivo} during transcription, nuclear remodeling, and confined migration - yet how mechanical response varies from locus to locus, and how it relates to epigenetic state, remains unclear. We develop a theory to infer locus-specific viscoelasticity from three-dimensional genome organization. Using chromatin structures derived from contact maps, we calculate frequency-dependent storage and loss moduli for individual loci and establish that the mechanical properties are determined both by chromatin epigenetic marks and organization. On large length scales, chromatin exhibits Rouse-like viscoelastic scaling, but this coarse behavior masks extensive heterogeneity at the single-locus level. Loci segregate into two mechanical subpopulations with distinct longest relaxation times: one characterized by single-timescale and another by multi-timescale relaxation. The multi-timescale loci are strongly enriched in active marks, and the longest relaxation time for individual loci correlates inversely with effective local stiffness. Pull–release simulations further predict a time-dependent susceptibility: H3K27ac-rich loci deform more under sustained forcing yet can resist brief, large impulses. At finer genomic scales, promoters, enhancers, and gene bodies emerge as ``viscoelastic islands'' aligned with their focal interactions. Together, these results suggest that chromatin viscoelasticity is an organized, epigenetically coupled property of the 3D genome, providing a mechanistic layer that may influence enhancer–promoter communication, condensate-mediated organization, and response to cellular mechanical stress. The prediction that locus-specific mechanics in chromatin are controlled by 3D structures as well as the epigenetic states is amenable to experimental test.
\end{abstract}

\maketitle
\end{CJK*}

\newpage

\section{Introduction}

The three-dimensional organization of chromatin plays a crucial role in gene regulation and genome function. Advances in chromosome conformation capture techniques, particularly Hi-C, have revealed the complex spatial architecture of the genome and its potential impact on function \cite{lieberman2009comprehensive,Dekker2024cell,Ruiz17COSYB,Vercellone25Physiology}. Significant progress has been made in elucidating the physical principles underlying genome organization~\cite{Dekker2024cell,Liu24ARB} and dynamics~\cite{Nozaki17MolCell, shiNatCommun2018,maeshimaColdSpringHarbPerspectBiol2021,salariGenomeRes2022,Thirumalai25ARPC, mazzocca2025}.  Recently, real-time dynamics of chromatin loci, with a focus on enhancer-promoter interactions,  have been studied using experiments~\cite{brucknerScience2023} and theory~\cite{shiStaticThreedimensionalStructures2025}. However, the mechanical properties that govern this organization remain comparatively less explored. Nuclear and chromatin mechanics play a key role in cellular mechanotransduction \cite{kalukulaMechanicsFunctionalConsequences2022,dupontMechanicalRegulationChromatin2022}, influencing processes such as cell migration, during which the nucleus and chromatin could deform under mechanical stresses \cite{davidsonDesignMicrofluidicDevice2015, jacobsonMigrationSmallPore2018}, potentially leading to DNA damage \cite{pfeifer2019nuclear, shahNuclearDeformationCauses2021}, genome reorganization \cite{Amar2021aplbioeng, golloshiConstrictedMigrationAssociated2022}, alterations in epigenetic states \cite{hsiaConfinedMigrationInduces2022, songTransientNuclearDeformation2022}, and transcriptional regulation \cite{Yoo2025}.

Recent advances have reshaped our understanding of the nucleus as a viscoelastic, hierarchically organized material whose mechanics emerge from dynamic chromatin crosslinking, polymer entanglement, and large-scale nuclear architecture \cite{dupontMechanicalRegulationChromatin2022, rodenburgHeterogeneityFeatureUnraveling2025, coulonInterphaseChromatinBiophysics2025}. Live-cell micromanipulation experiments have shown that human interphase chromatin behaves as a soft, weakly entangled polymer network exhibiting Rouse-like viscoelastic behavior \cite{Keizer2022science}. At the whole nucleus level, live-cell imaging and displacement correlation spectroscopy suggest that chromatin undergoes a local sol-gel transition during stem cell differentiation, transforming from a homogeneous Maxwell-like fluid into a composite material with coexisting solid-like heterochromatin and fluid-like euchromatin phases \cite{eshghiInterphaseChromatinUndergoes2021}. Perturbations in chromatin crosslinkers have been found to substantially alter nuclear stiffness. For instance, depletion of HP1$\alpha$ reduces both nuclear and mitotic chromosome rigidity, highlighting the role of protein-mediated crosslinking in maintaining elastic integrity of the nucleus \cite{stromHP1aChromatinCrosslinker2021}. On the other hand, simulations suggest that such crosslinking contributes to macroscopic elasticity only when heterochromatin is tethered to the nuclear lamina, forming a peripherally anchored, gel-like network that transmits mechanical stress \cite{attarPeripheralHeterochromatinTethering2025}. It has been shown that on the DNA scale, condensin complexes  act as transient crosslinkers that markedly increase both the viscosity and the elasticity of entangled DNA \textit{in vitro} \cite{confortoComplexRheologyCondensin2025}, revealing a possible ATP-independent mechanism by which Structural Maintenance Complexes modulate chromatin rheology. Complementary polymer simulations suggest that spatial variations in chromatin bending rigidity alone can drive segregation of euchromatin and heterochromatin, coupling mechanical heterogeneity to nuclear compartmentalization \cite{girardHeterogeneousFlexibilityCan2024}. Together, these findings support a physical picture in which chromatin mechanics arise from a spectrum of relaxation times governed by intra-chromosomal interactions, crosslinking or loop extrusion dynamics, local stiffness, and boundary tethering—allowing the nucleus to function as an adaptive viscoelastic material that integrates molecular activity with cellular-scale mechanical response.

Epigenetic modifications, particularly histone marks, regulate chromatin structure and function. How epigenetic states systematically tune locus-specific chromatin mechanics to control its function  is unclear. Existing perturbation studies that globally alter acetylation or methylation report concomitant changes in chromatin and nuclear mechanics \cite{stephensChromatinLaminDetermine2017, Strom24Cell}, supporting an association between epigenetics and material response. However, these measurements are largely qualitative and coarse-grained, leaving unresolved how mechanics vary from locus to locus and how that heterogeneity maps onto specific epigenetic states.   Our approach addresses this limitation by resolving chromatin mechanics at single-locus resolution and directly linking the heterogeneous dynamics   to epigenetic states. We computed the storage and loss moduli of chromatin loci by extending the HIPPS-DIMES framework, which has been used to determine the 3D chromatin structures from the experimental Hi-C contact maps \cite{Shi2021PRX, Shi2023NatComm,shiStaticThreedimensionalStructures2025}. 
We show that euchromatin and heterochromatin display distinct mechanical behavior. Regions enriched for H3K27ac (active promoters/enhancers) frequently exhibit multi-timescale relaxation dynamics in which the ratio of the loss modulus to the storage modulus ($G^{\prime\prime}/G^{\prime}$) crosses unity multiple times, indicating the presence of several characteristic timescales. 
By contrast, regions enriched for H3K9me3 or H3K27me3 show simpler behavior with a single dominant timescale. Notably, the longest relaxation time in the active regions tends to exceed that in the repressive regions, consistent with more restricted motion on intermediate timescales for transcription-active gene loci observed in experiments and models \cite{Shin2024pnas, Nozaki2023sciadv,Zidovska13PNAS}. These results support a functional link between epigenetic state and chromatin mechanics. More broadly, they suggest that epigenetic remodeling may regulate not only gene expression programs but also how specific genomic regions store and dissipate mechanical energy under perturbations. Finally, we discuss the potential relevance of these findings to the formation and function of transcriptional condensates \cite{waghCurrentOpinioninStructuralBiology2021, peiTranscriptionRegulationBiomolecular2025}. More broadly, our theory provides testable predictions for how perturbations that alter chromatin interactions or epigenetic marks should reshape locus-specific viscoelasticity, offering a quantitative route to connect genome regulation with chromatin mechanics.

\section{Results}
\subsection{Region-averaged viscoelastic properties in GM12878 cells}
\begin{figure*}[t]
\centering
\includegraphics[width=.8\linewidth]{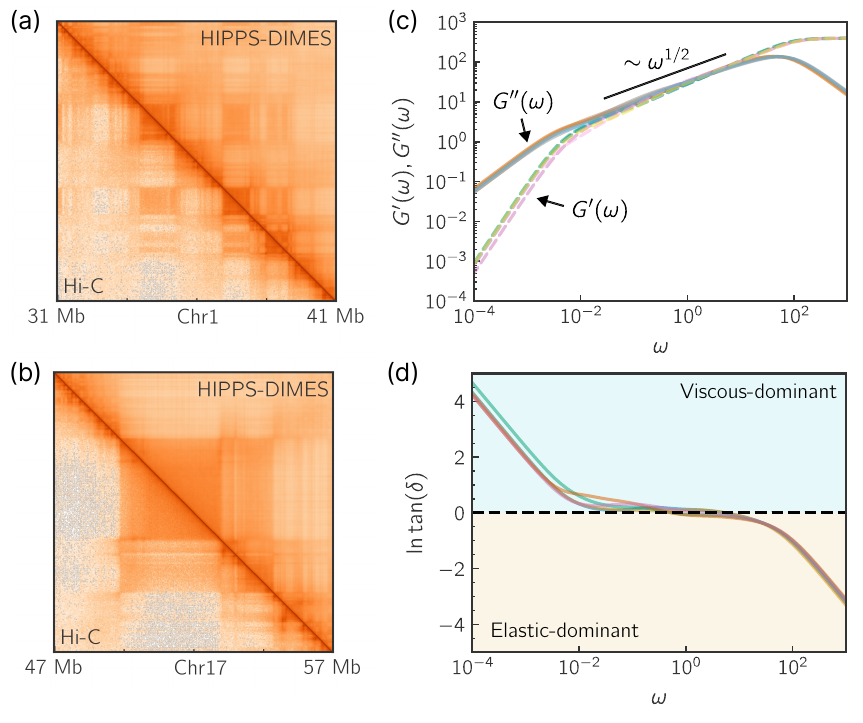}
\caption{\textsf{Region-averaged viscoelastic properties of chromosomes in GM12878. (a) Comparison of the contact map between Hi-C and HIPPS-DIMES for a 10 Mb region in chromosome 1. Contact maps are calculated directly from the 3D structures determined using the HIPPS-DIMES framework. (b) Same as (a) but for Chr17: 47 Mb to 57 Mb. (c) Storage ($G^{\prime}$) and loss modulus ($G^{\prime\prime}$) for GM12878. Solid lines are $G^{\prime\prime}(\omega)$ and dashed lines are $G^{\prime}(\omega)$. The colors represent different regions. All regions studied are 10 Mb long. Solid black line is a guide to the eye showing $\sim \omega^{1/2}$ dependence. (d) Logarithm of the ratio between loss modulus and storage modulus, $\tan(\delta) = G^{\prime\prime}(\omega)/G^{\prime}(\omega)$.}}
\label{fig:gm12878_modulus}
\end{figure*}

We first investigated the viscoelastic properties of chromatin at the 10 Mb scale. To this end, we selected five genomic regions from different chromosomes in the GM12878 cell line (see Methods for details), each spanning 10 Mb. As a validation step, we confirmed that the HIPPS-DIMES framework produces structural ensembles that are consistent with the Hi-C data. Figures~\ref{fig:gm12878_modulus}(a,b) show that the contact maps derived from the 3D structures determined using the HIPPS-DIMES method  are in excellent agreement with the experimental Hi-C contact maps (see Appendix A and Figure \ref{fig:gm12878_10mb_contact_maps_combine} for three additional regions investigated). 

We then computed the storage and loss moduli, $G^{\prime}$ and $G^{\prime\prime}$, for the five regions using Eq.~\ref{eq:eq2}. As shown in Figure~\ref{fig:gm12878_modulus}(c), the results from different chromosomes collapse onto one another, indicating that at the 10 Mb length scale, viscoelastic properties are relatively uniform across the genome.  We speculate that this homogenization arises because, at the $\sim$10 Mb scale, there are extended segments of both euchromatin and heterochromatin. As a result, the average mechanical response is similar across distinct regions.

Interestingly, in the intermediate frequency regime, both $G^{\prime}$ and $G^{\prime\prime}$ scale approximately as $\omega^{1/2}$, which is in accord with the predictions of the Rouse model \cite{doi1988theory}. This finding demonstrates that, on this scale, chromatin in GM12878 cells exhibits Rouse-like viscoelastic behavior, which is in agreement with  experimental observations \cite{Keizer2022science} probing the effect of mechanical force on a labeled locus. 

To further characterize the frequency-dependent response, we calculated the loss tangent, $\tan(\delta) = G^{\prime\prime}/G^{\prime}$. Physically, $\tan(\delta) > 1$ indicates that viscous contributions dominate over elasticity, whereas $\tan(\delta) < 1$ implies elastic dominance.  At low frequencies (long times) $\tan(\delta)$ exceeds unity (viscous effects dominate), implying that on the 10 Mb scale chromatin is fluid-like. Figure~\ref{fig:gm12878_modulus}(d) shows that $\ln \tan(\delta)$ remains near zero in the intermediate frequency range, again consistent with the Rouse  model predictions and highlighting the viscoelastic nature of chromatin. We also examined viscoelastic properties at the scale of entire chromosomes by applying the HIPPS-DIMES framework at 100 kb resolution contact maps. The results, shown in Appendix B and Figure \ref{fig:gm12878_chr3_combine}, exhibit similar scaling behavior and Rouse-like viscoelasticity, indicating the robustness of these findings.

\subsection{Locus-specific viscoelasticity}

\begin{figure*}[t]
\centering
\includegraphics[width=\textwidth]{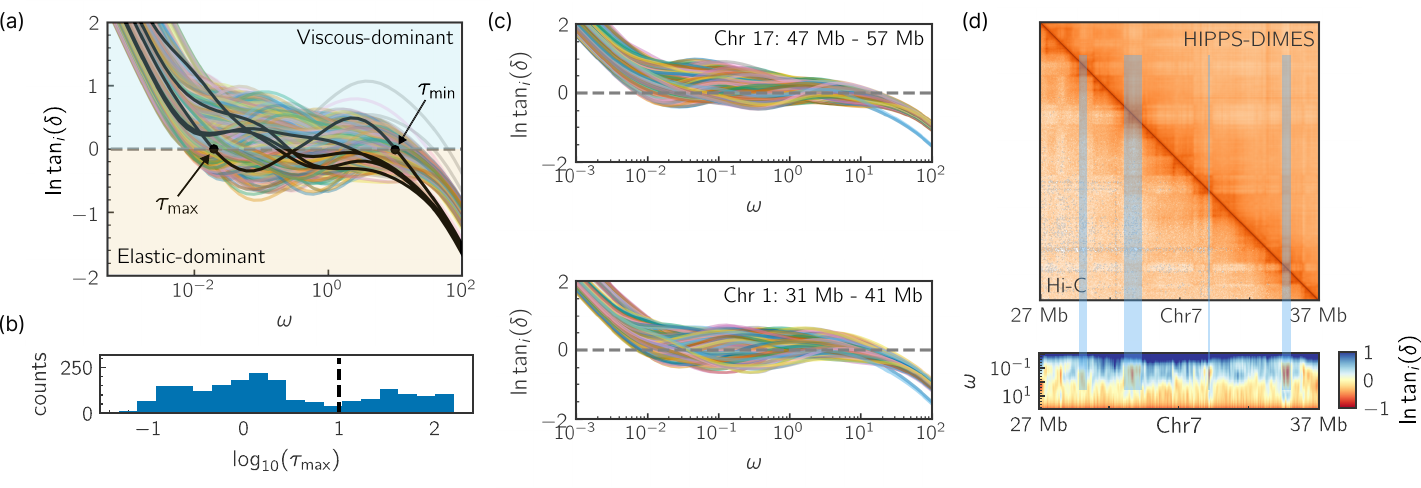}
\caption{\textsf{Locus-specific viscoelasticity  in GM12878 chromosomes. (a) Frequency dependence of $\tan_{i}(\delta)$ for individual loci from all 5 regions investigated, with each curve representing a single locus. Several representative loci are highlighted in black. Distinct crossing behavior at $\tan_{i}(\delta)=1$ reveals variations in the relaxation dynamics across the loci. For selected loci, arrows indicate $\omega_{\min}$ and $\omega_{\max}$, the lowest- and highest-frequency crossings of $\tan_i(\delta)=1$, and the corresponding relaxation timescales $\tau_{\max}=1/\omega_{\min}$ and $\tau_{\min}=1/\omega_{\max}$.
(b) Histogram of the largest relaxation time, $\tau_{\text{max}} = 1/\omega_{\text{min}}$, showing a bimodal distribution that points to two distinct subpopulations of loci with different viscoelasticity. Vertical dashed line marks the separation of two subpopulations.
(c) $\ln \tan_i(\delta)$ as a function of $\omega$ for Chr 17: 47 Mb - 57 Mb and Chr1: 31 Mb - 41 Mb.
(d) 
Hi-C contact map (top) and the corresponding heatmap of $\ln\tan_i(\delta)$ (bottom) for the same genomic region, with loci aligned along the horizontal axis. In the $\ln\tan_i(\delta)$ panel, the vertical axis denotes $\omega$ (increasing from top to bottom), and the colormap encodes the sign of $\ln\tan_i(\delta)$ (blue: $G''_i > G'_i$, white: $G''_i = G'_i$, orange/red: $G'_i > G''_i$). Shaded boxes highlight examples where ``islands’’ of complex viscoelastic behavior coincide with domain-like structures in the contact map.}}
\label{fig:gm12878_single_loci_modulus_combine}
\end{figure*}

Although the region-averaged moduli demonstrate that chromatin exhibits Rouse-like viscoelastic behavior, it is likely that potential heterogeneity manifests itself only at the level of individual loci. To reveal structural heterogeneity~\cite{Finn19Cell}, we calculated the locus-specific storage and loss moduli using Eq.~\ref{eq:eq3}.
The locus-level analysis reveals substantial variations in the viscoelastic properties across the genomic regions. Figure~\ref{fig:gm12878_single_loci_modulus_combine}(a) shows the frequency dependence of $\tan_i(\delta)$ for individual loci, with each curve corresponding to a single locus across all five regions. Figure~\ref{fig:gm12878_single_loci_modulus_combine}(c) shows two specific regions as examples. Several representative loci are highlighted in black solid lines. Notably, $\ln\tan_i(\delta)$  crosses zero either once or multiple times, reflecting distinct relaxation behavior associated with various loci. Such non-monotonic behavior of $\tan_i(\delta)$ has also been observed recently in \textit{in vitro}–reconstituted chromatin condensates formed from nucleosome arrays \cite{Zhou25Science}.

To quantify the distinct locus-dependent relaxation behavior in Figure~\ref{fig:gm12878_single_loci_modulus_combine}(a), we define the largest relaxation time as $\tau_{\text{max}} = 1/\omega_{\text{min}}$ where $\omega_{\text{min}}$ is the smallest frequency at which $\tan(\delta) = 1$. The distribution of $\tau_{\text{max}}$ across the loci (Figure~\ref{fig:gm12878_single_loci_modulus_combine}(b)) exhibits a bimodal structure, indicating two distinct subpopulations of mechanical behavior.
We next asked whether the locus-to-locus heterogeneity in the viscoelastic response is related to the underlying chromatin organization. To address this, we align the Hi-C contact map with a heatmap of $\ln\tan_i(\delta)$ for the same region (Fig.~\ref{fig:gm12878_single_loci_modulus_combine}(d)). In the $\ln\tan_i(\delta)$ heatmap, the vertical axis is the driving frequency $\omega$ (increasing from top to bottom), and the color encodes the sign of $\ln\tan_i(\delta)$: blue for $\ln\tan_i(\delta) > 0$ ($G''_i > G'_i$), white for $\ln\tan_i(\delta) = 0$ ($\tan_i(\delta)=1$), and orange/red for $\ln\tan_i(\delta) < 0$ ($G'_i > G''_i$). Loci with a simple, monotonic decay of $\tan_i(\delta)$ exhibit a single blue–white–orange transition as $\omega$ increases. In contrast,  loci that display non-monotonic behavior and three crossings characterized by $\tan_i(\delta)=1$ display alternating bands of blue and orange along the frequency axis. These complex patterns are not scattered randomly along the genome, but instead form contiguous ``islands’’ in the heatmap. Notably, many of these islands coincide with domain-like structures (Topologically Associating Domains or contact domains) in the contact map, as highlighted by the shaded boxes in Fig.~\ref{fig:gm12878_single_loci_modulus_combine}(d). This result  indicates that viscoelastic heterogeneity is organized at the domain scale and is reflected in the local 3D chromatin architecture. In other words, the dynamical response can be gleaned from the chromatin organization.

\subsection{Heterogeneity in locus-specific viscoelasticity is associated with histone modifications}
Figure~\ref{fig:gm12878_single_loci_modulus_combine}(b) shows that the distribution of the largest relaxation times of individual loci roughly follows a bimodal distribution. Combined with the visual correspondence between the Hi-C contact domains and patterns of viscoelastic behavior (Figure \ref{fig:gm12878_single_loci_modulus_combine}(d)), this key finding suggests that chromatin mechanical heterogeneity may be biologically regulated. We hypothesize that loci marked by different histone modifications should exhibit distinct viscoelastic behavior. To test this notion, we examined six histone modification ChIP-seq datasets: H3K27ac, H3K4me1, H3K4me3, H3K36me3, H3K27me3, and H3K9me3. Among these, H3K27ac, H3K4me1, H3K4me3, and H3K36me3 are euchromatin marks with distinct biological roles. For example, H3K27ac is associated with active promoters and enhancers, while H3K36me3 marks actively transcribed gene bodies. In contrast, H3K27me3 and H3K9me3, associated with heterochromatin, are repressive marks. In particular, H3K27me3 is enriched at Polycomb-repressed domains.

\begin{figure*}[t]
\centering
\includegraphics[width=\linewidth]{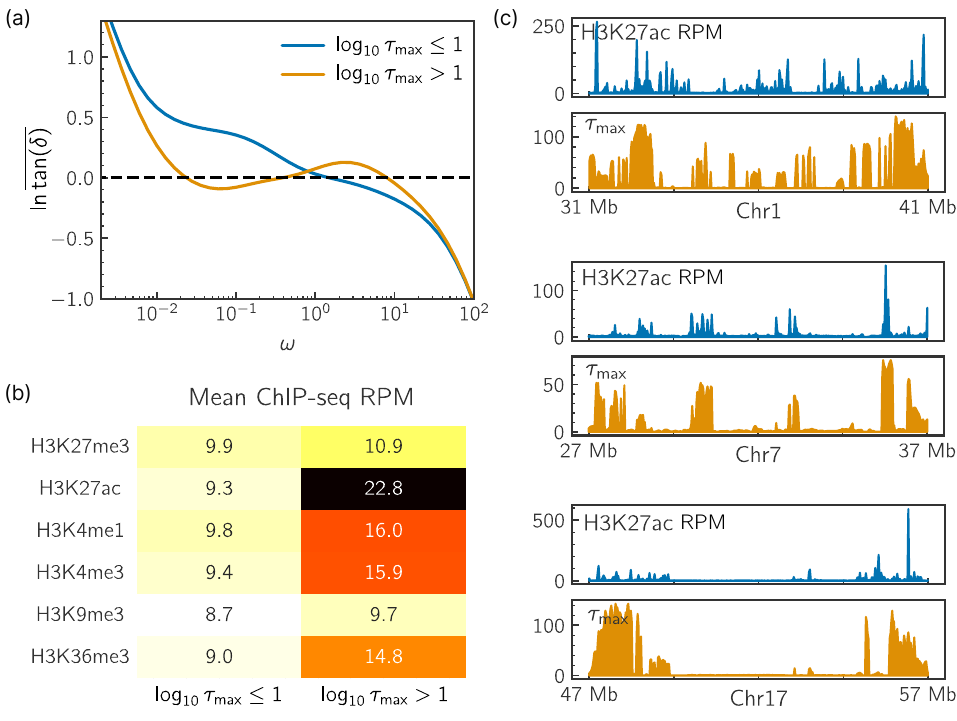}
\caption{\textsf{Association between chromatin viscoelastic properties and histone modifications in GM12878 cells.
(a) Average $\ln\overline{\tan(\delta)}$ profiles for loci grouped by their largest relaxation time, $\tau_{\text{max}}$. Loci with $\log_{10}\tau_{\text{max}} > 1$ (long relaxation times) exhibit complex behavior with three zero crossings, whereas loci with $\log_{10}\tau_{\text{max}} \leq 1$ (short relaxation times) show a simpler profile with a single crossing.
(b) Mean histone modification ChIP-seq read counts per million (RPM) across the two groups. Loci with long relaxation times are enriched for active chromatin marks (H3K27ac, H3K4me1, H3K4me3, H3K36me3), while loci with short relaxation times are relatively enriched in repressive marks (H3K27me3, H3K9me3).
(c) Representative genomic regions from chromosomes 1, 7, and 17 illustrating the correspondence between high H3K27ac RPM and loci with long relaxation times $\tau_{\text{max}}$.}}
\label{fig:viscoelastic_epigenetics_combine}
\end{figure*}

Based on the positions of the two peaks in the distribution of $\tau_{\text{max}}$ (Figure~\ref{fig:gm12878_single_loci_modulus_combine}(b)), we used $\log_{10}\tau_{\text{max}} = 1$ as a threshold to divide the loci into two groups: $\log_{10}\tau_{\text{max}} \leq 1$ (short relaxation times) and $\log_{10}\tau_{\text{max}} > 1$ (long relaxation times). We then compared the viscoelastic behavior of these two groups. As shown in Figure~\ref{fig:viscoelastic_epigenetics_combine}(a), the loci with long relaxation times exhibit a complex profile, with $\ln\tan_i(\delta)$ crossing zero three times. In contrast, loci with short relaxation times display a simpler profile, crossing zero only once, consistent with a single characteristic relaxation timescale. Interestingly, although the long-relaxing loci have a larger $\tau_{\text{max}}$, they also exhibit smaller $\tau_{\text{min}} = 1/\omega_{\text{max}}$ than the single relaxation time of the short-relaxing group. Physically, this implies that these loci behave fluid-like at short timescales, transitioning to a more solid-like state at intermediate timescales. At long times there is a transition to a fluid-like state, reflecting the inherently viscoelastic and ultimately non-solid nature of chromatin.

We next wondered whether these two mechanical subpopulations correspond to differences in histone modifications. Figure~\ref{fig:viscoelastic_epigenetics_combine}(b) shows that loci with $\log_{10}\tau_{\text{max}} > 1$ are enriched in active histone marks (H3K27ac, H3K4me1, H3K4me3, H3K36me3) compared to loci enriched in repressive marks (H3K27me3 and H3K9me3).
Notably, H3K27ac levels in the long-relaxing loci are at least twice as high as those in the short-relaxing loci. The strong association can be directly visualized by comparing histone modification tracks with relaxation times. Three representative examples from chromosomes 1, 7, and 17 shown in Figure~\ref{fig:viscoelastic_epigenetics_combine}(c) confirm that regions with high H3K27ac signal often coincide with loci exhibiting long relaxation times.
We also examined the association between viscoelastic properties and histone modifications at the scale of entire chromosomes (Appendix B). The results, shown in Figure~\ref{fig:gm12878_chr3_combine}, exhibit similar patterns of enrichment. 

\subsection{Structural basis of locus-specific viscoelasticity}
Next, we asked whether loci with distinct viscoelastic behavior also differ in their structural organization. The fractal dimension of chromatin has been experimentally measured and proposed as an important parameter linking 3D genome architecture to function \cite{HuangSciAdv2020, AlmassalhaSciAdv2025}. To quantify this, we computed the fractal dimension, $D_f$, from our model. For each 10 Mb region, we generated 10,000 conformations and calculated the ensemble-averaged number of neighbors, $\langle n\rangle$, for each locus as a function of spatial cutoff distance $r$. The scaling relation $\langle n\rangle \propto r^{D_f}$ was fit in the range $0.1 < r < 1$ to obtain fractal dimension, $D_f$, analogous to a mass fractal dimension (details are in the Methods).

\begin{figure*}[t]
\centering
\includegraphics[width=\linewidth]{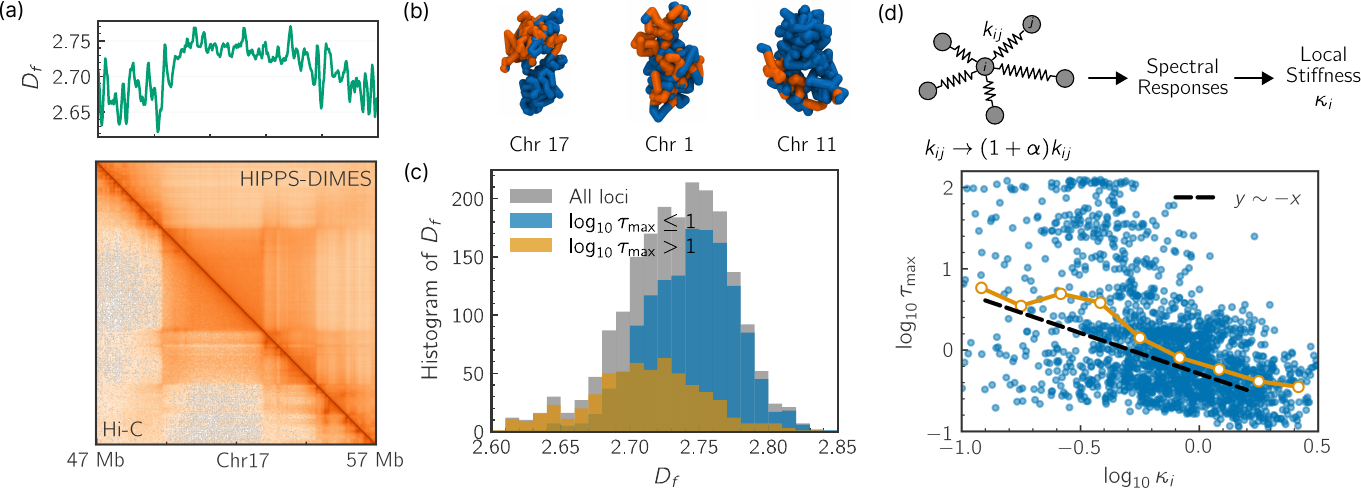}
\caption{\textsf{Structural organization of slow- and fast-relaxing loci in GM12878 cells.
(a) Locus-specific fractal dimension $D_f$ for chromosome 17 (47–57 Mb), computed from the scaling of the number of neighbors with distance.
(b) Representative 3D structures at 100 kb resolution sampled from the HIPPS-DIMES ensemble, with loci colored by relaxation group: slow-relaxing loci (orange) and fast-relaxing loci (blue). The slow-relaxing loci form spatially loose but distinct clusters. In the structure, each 100 kb segment is represented by a single 3D coordinate, and the structures are visualized using the licorice representation.
(c) Distributions of $D_f$ for slow and fast-relaxing loci are broad, reflecting the variations in the local environment. Slow-relaxing loci exhibit lower average fractal dimension ($\overline{D_f} = 2.71$) compared to fast-relaxing loci ($\overline{D_f} = 2.75$), indicating that active loci are organized in looser, less compact structures, whereas repressive loci are more structurally compact.
(d) Longest relaxation time $\tau_{\max}$ for each locus plotted against the locus-specific local stiffness $\kappa_i$ (scatter). The orange curve shows the binned mean. The black dashed line is a guide to the eye with slope $-1$. $\kappa_i$ is computed from the connectivity-spectrum response to a perturbation that rescales all the couplings $k_{ij}$ associated with locus $i$.}
}
\label{fig:gm12878_modulus_structure_combine}
\end{figure*}

Figure~\ref{fig:gm12878_modulus_structure_combine}(a) shows that the locus-specific $D_f$ profile for chromosome 17 (47–57 Mb) has considerable variation. To visualize the structural basis of the heterogeneity in $D_f$, we randomly sampled the 3D conformations and colored the loci according to their relaxation group: slow-relaxing loci in yellow and fast-relaxing loci in blue. Figure~\ref{fig:gm12878_modulus_structure_combine}(b) shows that the two groups are localized in spatially distinct clusters, consistent with the transcriptional hub model \cite{limCurrentOpinioninGenetics&Development2021}, where multiple enhancers and promoters colocalize beyond simple pairwise interactions. This observation aligns with our earlier finding (Figure~\ref{fig:viscoelastic_epigenetics_combine}(b)) that slow-relaxing loci are enriched in active histone marks.

We compared the structural properties of the two groups. Figure~\ref{fig:gm12878_modulus_structure_combine}(c) shows the distributions of $D_f$ for the slow- and fast-relaxing loci. On average, the fractal dimension of the slow-relaxing loci is lower, $\overline{D_f} = 2.71$, compared to $\overline{D_f} = 2.75$ for the fast-relaxing group. This suggests that active, slow-relaxing loci are organized in looser, less compact domains, whereas repressed, fast-relaxing loci are structurally more compact. Although the difference in the mean values is small, the distributions of $D_f$ are quite different (Figure~\ref{fig:gm12878_modulus_structure_combine}(c)), which again is a reflection of locus-specific heterogeneity.

To connect the relaxation dynamics to the local mechanical environment, we computed locus-specific local stiffness $\kappa_i$~\cite{vuSci.Adv.2022}. Briefly, $\kappa_i$ quantifies how strongly a small local stiffening of locus $i$ shifts the global relaxation spectrum. This is implemented by rescaling all the couplings, $k_{ij}$ associated with the $i^{th}$ locus by $(1+\alpha)$ taking the $\alpha\to 0 $ limit (i.e., evaluating the derivative of the relaxation eigenvalues with respect to $\alpha$) and normalizing  by the locus's thermal positional fluctuation (see Methods). Figure~\ref{fig:gm12878_modulus_structure_combine}(d) shows a scatter plot of $\log_{10}\tau_{\max}$ versus $\log_{10}\kappa_i$ for individual loci. The relationship is highly heterogeneous, with substantial locus-to-locus scatter spanning a broad range of both $\tau_{\max}$ and $\kappa_i$. To assess the overall trend, we computed the binned mean of $\log_{10}\tau_{\max}$ as a function of $\log_{10}\kappa_i$, which reveals an approximately inverse dependence, $\tau_{\max}\sim \kappa_i^{-1}$. This trend is consistent with overdamped Brownian dynamics of a harmonic mode where $\tau=\gamma/k$ and the friction coefficient is $\gamma$ and $k$ is the spring constant. In this analogy, $\kappa_i$ serves as a measure of an effective local stiffness.

\subsection{Active loci are more susceptible to sustained forces but resist impulsive forces}
\begin{figure*}[t]
\centering
\includegraphics[width=\linewidth]{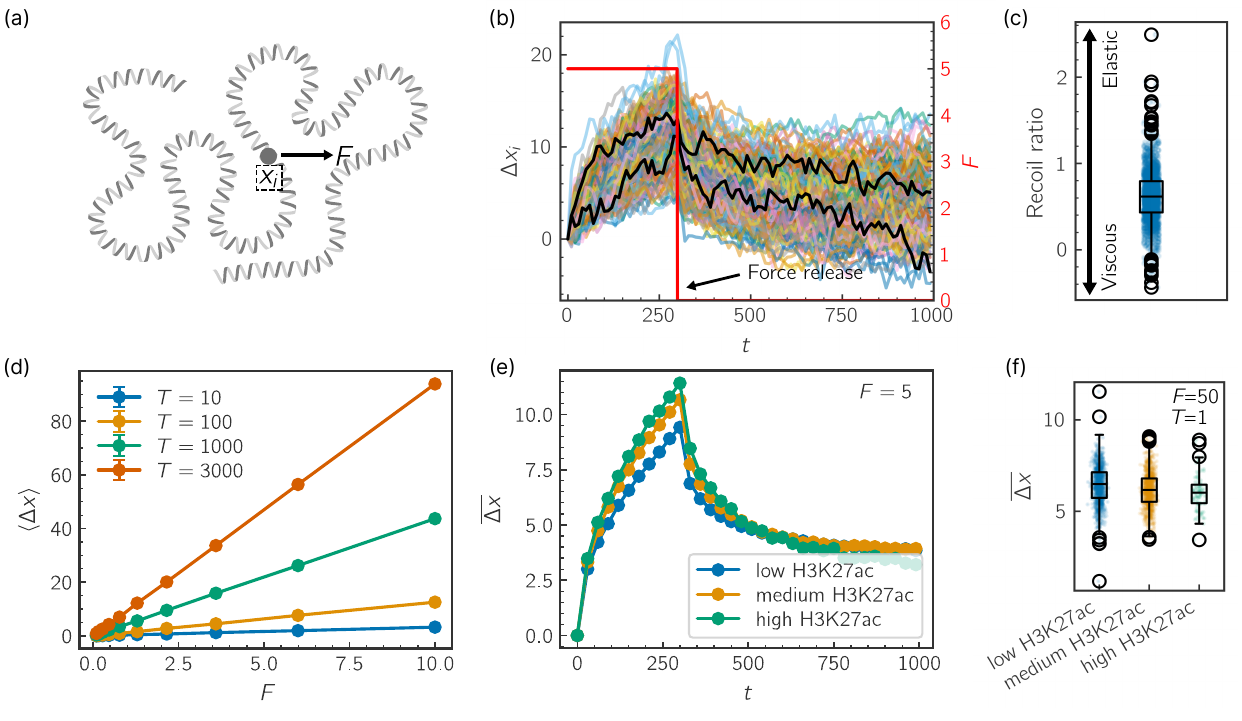}
\caption{\textsf{Force extension and recoil dynamics of genomic loci in GM12878 cells.
(a) Schematic showing application of mechanical  force $F$ on an individual locus labeled by the position $x_i$.  
(b) Displacement $\Delta x_i$ of individual loci as a function of time at $F=5$ directed along the positive $x$-axis (released at $t=300$, with relaxation followed until $t=1000$). Each curve represents a single locus (the black trajectories are shown as examples), with the substantial variations illustrating heterogeneous mechanical responses. The recoil dynamics is surprisingly slow.
(c) Box plot for recoil ratio   (defined as $(\Delta x_i(t=300) - \Delta x_i(t=1000)) / \Delta x_i(t=300)$) of individual loci. 
(d) Trajectory-averaged final displacement $\langle \Delta x\rangle$ for locus $i=50$ (32.25 Mb on Chr 1) at the end of force application as a function of force amplitude $F$. The force is applied for a duration $T$. 
$\langle \cdot \rangle$ denotes average over multiple trajectories.
(e) Mean force–recoil dynamics for loci grouped by H3K27ac enrichment, $\overline{\Delta x}$, obtained by averaging over individual loci shown in (a). Loci in the top 5\% of H3K27ac signal display the largest displacement, followed by those in the 50–95th percentile, while loci in the bottom 50\% show the smallest displacement. 
(f) The locus-average displacement, $\overline{\Delta x}$, with $F=50$ applied for a duration $T=1$ grouped by  H3K27ac enrichment. $\overline{\Delta x} = 6.4,\ 6.2,\ 6.0$ for loci with low, medium, and high H3K27ac enrichment, respectively.}}
\label{fig:force_recoil_combine}
\end{figure*}

Recent advances in single-chromosome and live cell nuclear manipulation experiments have provided quantitative insights into the response of chromatin and chromosomes to external mechanical forces~\cite{Keizer2022science,meijeringNonlinearMechanicsHuman2022, wittIonmediatedCondensationControls2024, bergamaschiHeterogeneousForceResponse2024,Strom24Cell}. To explain some of the findings in \cite{Keizer2022science,Strom24Cell}, we examined how individual genomic loci respond to mechanical stresses (Figure~\ref{fig:force_recoil_combine}(a)). To this end, we applied a localized mechanical force to selected loci and monitored both their displacement during force application and the recoil dynamics after force release (see Methods for details). This protocol is similar to that used in experiments, which were performed  in live-cell settings \cite{Keizer2022science}. Figure~\ref{fig:force_recoil_combine}(b) shows the locus-specific displacement $\Delta x_i$ (measured along the direction of the applied force) as a function of time. The magnitude of the pulling force used is $|F|=5$. The force  was applied for $t=300$, after which the system was allowed to relax up to $t=1,000$. Each curve in Figure~\ref{fig:force_recoil_combine}(b) represents the trajectory of a single locus. The responses are highly heterogeneous, consistent with the variations  found in $G^{\prime}$ and $G^{\prime\prime}$ (Fig. \ref{fig:gm12878_single_loci_modulus_combine}). Not surprisingly, upon force application, the loci move in the direction of the applied force; upon release of the force, they recoil, although in many cases the loci do not return fully to their original positions on the specified time scale.

To quantify the locus-dependent recoil dynamics,  we computed a dimensionless recoil ratio, defined as $(\Delta x_i(t=300) - \Delta x_i(t=1,000) )/ \Delta x_i(t=300)$. Smaller recoil ratio indicates predominantly viscous recoil (limited recovery) and larger recoil ratio implies more elastic recoil (greater recovery). Figure~\ref{fig:force_recoil_combine}(c) summarizes the distribution of recoil ratio as a box plot, revealing a broad spread across loci that spans viscous-like to elastic-like recoil dynamics. The substantial variations obtained theoretically are similar to the spread noted in experiments~\cite{Strom24Cell}.

As expected, stronger forces produced larger displacements. To quantify this, we measured the trajectory-averaged final displacement $\langle \Delta x \rangle$  at the end of the duration of the force-application  as a function of the force amplitude $F$ applied for a duration $T$. Figure~\ref{fig:force_recoil_combine}(d) shows that $\langle \Delta x \rangle$ scales linearly with $F$ with the slope that depends on $T$, thus confirming the expected applicability of the Rouse model in describing global response of chromatin to force~\cite{Keizer2022science,Strom24Cell}. 

Although the theory is formulated in reduced units, we can make a rough estimate of the physical scale of the force $F$ used in the simulations. Each locus represents 25 kb. Taking the effective size of a locus, $a$, to be in the range $70\ \text{nm} \text{–} 200\ \text{nm}$ \cite{shiNatCommun2018, Shi2021PRX}, the force scale ($F$) in the model corresponds to $k_B T/a  \sim 0.02\ \text{pN} \text{–} 0.06\ \text{pN}$ ($k_BT$ is taken to be $4.11\ \text{pN}\cdot\text{nm}$ at temperature of 298 $\text{K}$). Thus, $F = 5$ is approximately $0.1\ \text{pN} \text{–} 0.3\ \text{pN}$, which is within the range accessed in experiments \cite{Keizer2022science,Strom24Cell}. 
The slope of force-displacement (Figure~\ref{fig:force_recoil_combine}(d)) depends on the duration of application of force $T$, which implies that the protocol used in the stretch-release cycle will affect the displacement and the recoil dynamics as a function of time, as shown in the context of protein-protein interactions~\cite{Barsegov05PRL,Barsegov06BJ}. We estimate the slope to be in the range of $0.3\ \mu \text{m}/\text{pN}$ to $3\ \mu \text{m}/\text{pN}$ for $T=10$, $1\ \mu \text{m}/\text{pN}$ to $10\ \mu \text{m}/\text{pN}$ for $T=100$, $5\ \mu \text{m}/\text{pN}$ to $40\ \mu \text{m}/\text{pN}$ for $T=1,000$ and $10\ \mu \text{m}/\text{pN}$ to $90\ \mu \text{m}/\text{pN}$ for $T=3,000$. The experimental value of $\langle \Delta x\rangle/F$ for 5 minutes of force application is about $3\ \mu \text{m}/\text{pN}$~\cite{Keizer2022science}, which is in the range found theoretically for $T=100$.

Given that epigenetic states are associated with distinct viscoelastic responses (Figure~\ref{fig:viscoelastic_epigenetics_combine}(c)), we next asked whether loci with different histone modifications exhibit variations in the pull–recoil dynamics.  For this purpose, we first classified the loci into three groups based on their H3K27ac ChIP-seq signal: the lower 50th percentile, the 50th–95th percentile, and the top 5th percentile. For each group, we calculated the average force–recoil dynamics. As shown in Figure~\ref{fig:force_recoil_combine}(e), loci enriched in H3K27ac exhibited greater displacement during force application compared to loci with lower signals. The extent of displacement increases with H3K27ac levels when $F=5$ applied for $T=300$. This accords with the expectation that high (low) levels of H3K27ac enrichment  is associated with  less compact euchromatin relative to heterochromatin.  Because heterochromatin is denser than euchromatin, it implies that both compactness and epigenetic state control the heterogeneous response to force. Based on the finding that H3K27ac-enriched loci exhibit predominantly elastic responses relative to low-H3K27ac loci at short times ($t\approx 1$; Figure~\ref{fig:viscoelastic_epigenetics_combine}(a)), we anticipated that the force response should depend on both the forcing duration and the force magnitude. In other words, the loading rate ($\approx F/T$) determines the force response~\cite{Evans97BJ}. In accord with this expectation, we find that application of a large force for a brief time ($F=50$ for $T=1$) reverses the trend: on an average, loci with the highest H3K27ac enrichment exhibit the smallest displacement (Figure~\ref{fig:force_recoil_combine}(f)). These results suggest that active genomic regions are mechanically more susceptible to displacement only under sustained application of force, a finding that is consistent with recent experiments showing that histone acetylation softens the nucleus \cite{stephensChromatinLaminDetermine2017, shimamotoNucleosomeNucleosomeInteractions2017, stephensChromatinHistoneModifications2018}.  However, they are comparatively more resistant to rapid, short-timescale forcing. Our prediction may be tested in experiments by varying the duration of the stretch-release cycle, as done in a recent experiment~\cite{Keizer2022science}. 

The dual loading-rate dependent response  to mechanical force on active loci may have implications for transcriptional regulation. In particular, both RNA polymerase II translocation \cite{Shin2024pnas} and loop extrusion by the scrunching mechanism \cite{Takaki2021cellreports} result in physical forces being imparted to chromatin.  Our predictions raise the possibility the deformation of the active loci in such processes depends on both the amplitude of applied forces and the timescales over which they act.

There are striking similarities between the theory and the experimental findings~\cite{Keizer2022science,Strom24Cell}. First, the mean displacement varies linearly with $F$. More importantly, the calculated value of $\text{d} \langle \Delta x \rangle / \text{d} F$ is close to estimate made using the data presented in experiments (see Figure 2C in \cite{Keizer2022science}). Second, our calculations show that, although globally chromatin behaves like a Rouse polymer, there is considerable heterogeneity at the single locus level at all times. This supports the general conclusions made in the above cited experimental studies. Third, the recoil dynamics of the loci after applying force for a finite duration is slow and spread over a broad range of times (Figure~\ref{fig:force_recoil_combine}b. Similar findings have been observed in experiments~\cite{Strom24Cell}. The prediction that the recoil dynamics should depend on the loading rate (rate of force application), as shown in Figure~\ref{fig:force_recoil_combine} (d) -(f), awaits future experimental verification.

\subsection{Enhancers, promoters and gene regions display distinct complex viscoelasticity}

The results presented above were obtained using Hi-C contact maps at 25 kb resolution for chromosomes from GM12878 cells. Each locus corresponds to a 25,000 bp segment. Although we established strong associations between active histone marks and locus-dependent viscoelasticity, it is unclear whether fine-scale regulatory elements such as promoters, enhancers, and gene bodies exhibit similar variations at high resolution.
To provide a fine-grained picture of the dynamics at the nucleosome, we took advantage of a recent study~\cite{GoelNatGenet2023} that has produced high-resolution contact maps.  The Region Capture Micro-C (RCMC) data from mouse embryonic stem cells (mESC) \cite{GoelNatGenet2023} have generated patterns of interactions at high resolution, allowing us to examine the viscoelastic properties at near-nucleosome scale. 

We applied the HIPPS-DIMES framework to a locus spanning 0.14 Mb (Chr8, 85.68-85.82 Mb) at 250 bp resolution. This region contains multiple genes including JUNB and PRDX2. The calculated locus-specific loss tangent, $\tan_i(\delta)$, in Figure~\ref{fig:mESC_KLF1_combine}(a) shows that, just as in GM12878 cells, the loci in this region at 250-bp resolution also exhibit heterogeneous behavior.  The frequency dependence of several $\ln \tan_i(\delta)$ profiles are non-monotonic and cross $\ln \tan_i(\delta) = 0$ three times, whereas others show a simpler response. Two representative examples are highlighted in black.

Figure~\ref{fig:mESC_KLF1_combine}(b) compares the experimental contact and the calculated contact maps using the 3D coordinates generated by the HIPPS-DIMES method. The contact map reveals numerous nested looping (focal) interactions, marked by characteristic "looping"-type dots, as indicated by the arrows and circles. These interactions are primarily formed by promoters, enhancers, and gene bodies enriched in active histone modifications \cite{GoelNatGenet2023}. Such fingerprints are also reflected in the elevated H3K27ac level (see panel below the contact map in Figure~\ref{fig:mESC_KLF1_combine}(b)). The 250 bp resolution allows for precise identification of these regulatory elements. The prevalence of heterogeneity in the viscoelastic properties at all scales~\cite{Di16PNAS} (nucleosome level to megabase) shows that chromatin organization is self-similar.

The corresponding $\ln \tan_i(\delta)$ heatmap, shown below the contact map in Figure~\ref{fig:mESC_KLF1_combine}(b), reveals that regions forming nested looping (focal) interactions exhibit distinct ``viscoelastic islands''. These regions display complex behavior where $\tan_i(\delta)$ crosses unity three times, with the first crossing occurring at much higher frequencies compared to the surrounding chromatin. This indicates that these regulatory regions become viscous-dominant, suggesting enhanced dynamic behavior at short times. To quantify this behavior, we calculated $\omega_{\text{max}}$ for each 250 bp segment. The bottom panel of Figure~\ref{fig:mESC_KLF1_combine}(b) shows the $\omega_{\text{max}}$ track, with peaks marked by black arrows. These peaks mostly correspond to the positions of promoters, enhancers, and gene bodies that form the focal interactions observed in the contact map. These regulatory regions exhibit non-monotonic $\tan_i(\delta)$ behavior. The complex viscoelastic signature, predicted using only the contact maps, suggests that active regulatory elements possess multiple characteristic relaxation timescales.

Figure~\ref{fig:mESC_KLF1_combine}(c) provides a complementary, network-based view of the same region by representing the chromatin interaction network as a weighted locus–locus graph. For each locus pair, we first calculated  the ensemble-averaged interaction energy, $\langle \epsilon_{ij}\rangle = k_{ij} \langle r_{ij}^2\rangle$, and converted it into an edge with positive weight $w_{ij}=\exp(\langle \epsilon_{ij}\rangle)$. Thus, by construction, large $w_{ij}$ implies strong effective attractive association. We then used $w_{ij}$ both to define the weighted edges and to set the forces in a Fruchterman–Reingold force-directed layout~\cite{Fruchterman91Software}, yielding an intuitive “interaction-space” embedding of the region. Node sizes are scaled by H3K27ac RPM, and node colors encode the locus-specific values of $\omega_{\text{max}}$. Strikingly, loci with stronger H3K27ac signal (larger nodes) preferentially cluster together, consistent with the observation that they form focal interactions in the RCMC contact map. Moreover, this cluster is enriched in magenta-colored nodes (higher $\omega_{\text{max}}$), demonstrating the link between active regulatory elements, focal interaction hubs, and complex viscoelastic signatures. While Figure~\ref{fig:mESC_KLF1_combine}(c) emphasizes this abstract network organization in the interaction space, we next asked how the same loci are arranged in real space.
Figure~\ref{fig:mESC_KLF1_combine}(d) presents the calculated representative 3D structural snapshots of the region, with loci colored according to their $\omega_{\text{max}}$ values. These snapshots show that these regions are more open and exposed, and they do not always form compact clusters together.

These findings show that the complex viscoelastic behavior calculated for chromatin at the 25 kb scale in GM12878 cells is also recapitulated and refined at the nucleosome level in mESC cells. In other words, the heterogeneous behavior at the single locus level and the connection to the underlying chromatin organization may be universal across cell lines at all scales.  The enhanced resolution in RCMC reveals that the individual regulatory elements—promoters, enhancers, and gene bodies exhibit mechanical signatures that are distinct from the surrounding non-regulatory and non-gene-containing chromatin.

\begin{figure*}[t]
\centering
\includegraphics[width=\linewidth]{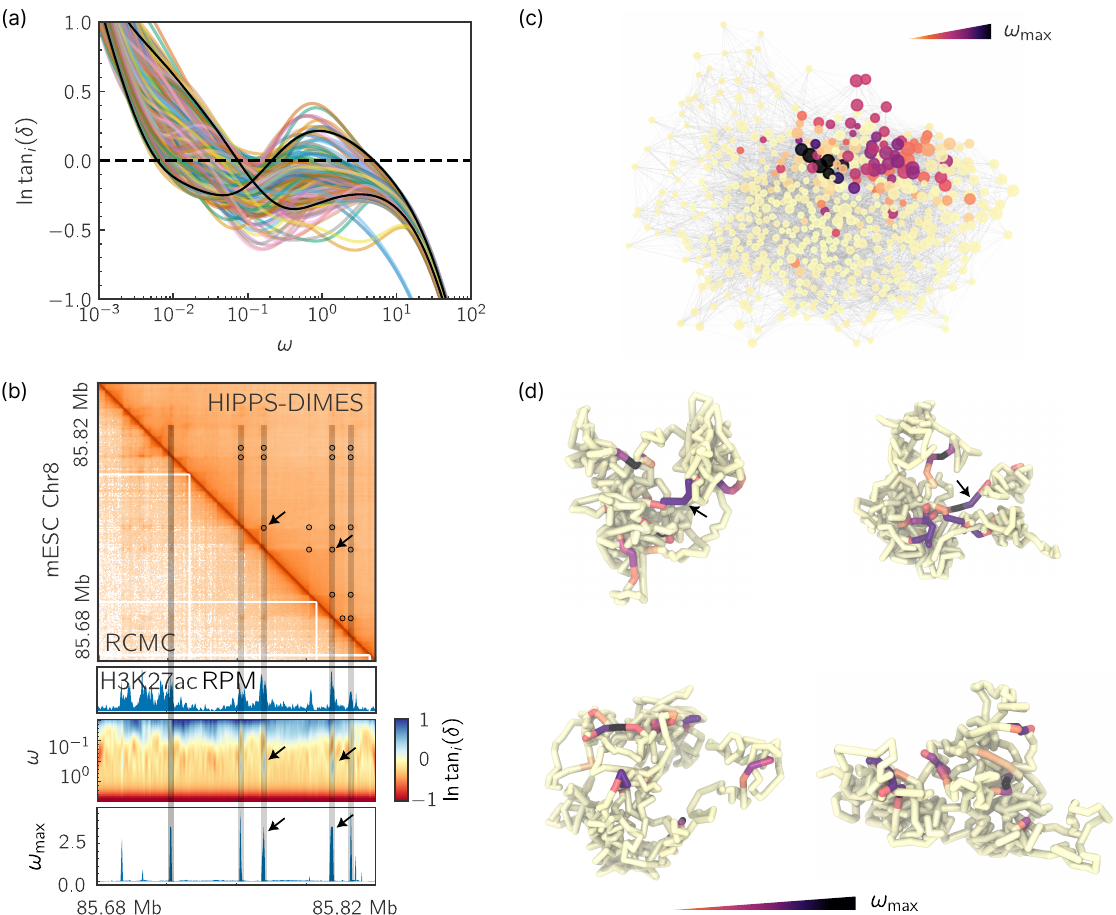}
\caption{\textsf{\label{fig:mESC_KLF1_combine}High-resolution viscoelastic properties of regulatory elements in mESC cells.
(a) Locus-specific loss tangent $\ln \tan_i(\delta)$ for mESC Chr8: 85.68 Mb - 85.82 Mb region. Each locus is 250 bp. Two examples of the dependence of $\ln \tan_i(\delta)$ on $\omega$ are shown in black.
(b) Comparison of experimental RCMC contact map with that calculated using 3D structures generated using HIPPS-DIMES for the Chr8: 85.68–85.82 Mb region. Black arrows show nested looping interactions formed by promoters, enhancers, and gene bodies (black circles). The $\ln \tan_i(\delta)$ heatmap shows distinct ``viscoelastic islands'' (black arrow) for regions forming these interactions. The  $\tan_i(\delta)$ profiles cross unity three times. The $\omega_{\text{max}}$ track (bottom panel) shows peaks (marked by black arrows) that correspond to regulatory elements forming focal interactions. Shaded boxes mark the correspondence between ``looping" interaction, H3K27ac enrichment, ``viscoelastic islands'' and peaks of $\omega_{\text{max}}.$
(c) Graph representation of the locus interaction network. Each node represents a locus. Node size is scaled by H3K27ac RPM, edge width by the exponential of the pairwise interaction energy, and node color by $\omega_{\text{max}}$. The network is visualized using a Fruchterman–Reingold force-directed layout.
(d) Representative 3D structural snapshots of the region at 250-bp resolution, with loci colored according to their $\omega_{\text{max}}$ values (magenta: high values, yellow: low values). The structures are visualized using the licorice representation. Arrows mark examples of regions with high $\omega_{\text{max}}$ values.}}
\label{fig:mESC_KLF1_combine}
\end{figure*}

\section*{Discussion}
We developed a theory that reveals a direct link between locus-specific viscoelasticity of chromatin and epigenetic states, including associated changes in the acetylation marks, thus providing new insights into the mechanical basis of genome organization and  function. Let us summarize the major findings. (1) At the 10 Mb to whole chromosome scale, chromatin dynamics exhibit Rouse-like viscoelastic behavior, with both storage and loss moduli scaling as $\omega^{1/2}$ in the intermediate frequency regime. This behavior reflects averaging over a diverse local mechanical environment across large genomic regions. (2) Importantly, at the single-locus level, there is substantial heterogeneity in the viscoelastic response, with relaxation times following a bimodal distribution that correlates well with epigenetic alterations. Within our theoretical framework, the origin of the heterogeneity can be traced to the variations in the spring constants that describe the inter-loci interactions. Recent experimental findings~\cite{bergamaschiHeterogeneousForceResponse2024} using force spectroscopy that probes the mechanical response of chromatin in an isolated nucleus emphasized that multiple spring constants are required to interpret chromatin dynamics, which was shown in an earlier theoretical study~\cite{Shi2023NatComm}.  (3) There is a striking association between histone modifications and viscoelastic behavior. Loci that are enriched in active chromatin marks (H3K27ac, H3K4me1, H3K4me3, H3K36me3) exhibit complex relaxation dynamics with multiple characteristic timescales. In sharp contrast, loci marked by repressive modifications (H3K27me3, H3K9me3) exhibit simpler behavior characterized by a single relaxation time. (4) Notably, active loci have longer relaxation times, consistent with more restricted motion on intermediate timescales. This  finding  aligns with recent computational studies~\cite{Shin2024pnas} and experimental observations of reduced mobility in transcriptionally active regions \cite{nagashimaSingleNucleosomeImaging2019, Nozaki2023sciadv,Zidovska13PNAS}. 

\textbf{Link to 3D structures:} The differences in the viscoelastic behavior between active and repressive loci are reflected in their 3D structural organization. Active loci exhibit lower fractal dimensions and form spatially distinct, less compact clusters, which supports  the essentials of the transcriptional hub model \cite{limCurrentOpinioninGenetics&Development2021}. In contrast, repressive loci are more structurally compact with higher fractal dimensions, suggesting that mechanical properties are intimately linked to chromatin folding.

A seemingly counter-intuitive result is that active loci, despite being less compact, tend to have longer relaxation times. This suggests that relaxation is not determined by compactness alone, but by how a locus is mechanically embedded in the interaction network. To probe this, we computed a locus-specific local stiffness by quantifying how a small local stiffening of couplings around the $i^{th}$ locus shifts the relaxation spectrum. Our analysis suggests an approximately inverse relationship between local stiffness and relaxation time, which is consistent with our numerical results and provides a mechanistic explanation for slow relaxation in open, active regions.

\textbf{Fluid-like versus solid-like behavior:} Whether chromatin has solid-like~\cite{strickfaden2017condensed,de07NanoLett} (an elastic material) characteristics or is fluid-like~\cite{Keizer2022science,Gibson23PNAS} (does not resist shear) on a mesoscale has been controversial. By inserting magnetic particles in HeLa cell and probing the response to an external magnetic field (a setup that is not that dissimilar to the one used more recently), a relatively high value for the  Young's modulus was reported ~\cite{de07NanoLett}. Although not explicitly stated, the non-value of the Young's modulus implies solid-like behavior of the chromatin fiber.  By using microscopy~\cite{strickfaden2017condensed} techniques it has been argued that nucleosomal arrays are solid-like under a range of conditions. In contrast, both \textit{in vitro} imaging experiments~\cite{Gibson23PNAS} and the response of external magnetic particles localized to a 4 Mb locus in live cells \cite{Keizer2022science} suggest that chromatin is fluid-like. Our findings, which seem to be supported by recent experiments~\cite{Strom24Cell,Zhou25Science} and previous arguments~\cite{Zidovska20Cell} suggest that chromatin should be pictured as a mosaic with substantial variations in the epigenetically controlled mechanical properties. Like in viscoelastic polymers, the relaxation times span a range of time scales~\cite{Di18PNAS}, which we have shown here may be predicted using their static 3D structures. The latter can be determined either by Hi-C methods or direct imaging. Most importantly, we predict that the locus-specific viscoelasticity is encoded in the epigenetic states.

\textbf{Response to force:} Beyond characterizing the passive viscoelasticity, we examined how individual genomic loci respond to external mechanical stress using pulling simulations. Strikingly, the loci respond heterogeneously to applied forces, with displacement scaling approximately linearly with the amplitude of the external force.  The extent of displacement of the loci depends on the duration of the application of the force. Interestingly, the slope of the mean extension-force relation is surprisingly similar to experiments~\cite{Keizer2022science}. This suggests that the mechanical response of interphase chromatin may be universal.

An important prediction in our work is that H3K27ac enrichment is associated with a timescale-dependent force response. Under sustained forcing, displacement increases with H3K27ac level, consistent with H3K27ac-rich euchromatin being less compact. However, because H3K27ac-enriched loci are more elastic at short timescales, we expected sensitivity to forcing duration and the amplitude of the exerted force. Indeed, large perturbation, exerted over a short duration, reverses the trend: the highest-H3K27ac loci show the smallest displacement. Thus, active regions are more deformable under sustained loads but more resistant to impulsive forcing, consistent with acetylation-dependent softening of nuclear mechanics \cite{stephensChromatinLaminDetermine2017, shimamotoNucleosomeNucleosomeInteractions2017, stephensChromatinHistoneModifications2018}. This could matter for transcription, since RNA polymerase II~\cite{HerbertAnnuRev08} and loop extrusion impose forces on distinct timescales \cite{Shin2024pnas,Takaki2021cellreports}.


\textbf{Chromatin at the near-nucleosome level:} To explore chromatin mechanics at finer scales, we extended our analysis beyond the 25 kb resolution. The 25 kb scale contact maps are too coarse to capture individual regulatory elements such as enhancers, promoters, and gene bodies. We  applied our framework to high-resolution Region Capture Micro-C (RCMC) data from mouse embryonic stem cells (mESC) at 250 bp resolution, approaching nucleosome level detail. 
At  near-nucleosome resolution, we again find strong locus-to-locus heterogeneity in $\tan_i(\delta)$, including both simple and multi-timescale responses, indicating that mechanical diversity is present at multiple length scales. Strikingly, regions that form nested looping interactions in the RCMC contact map---primarily promoters, enhancers, and gene bodies with elevated H3K27ac---appear as distinct ``viscoelastic islands'': they exhibit three crossings of $\tan_i(\delta)=1$, with the first crossing shifted to higher frequencies, suggesting enhanced dissipation at short timescales. Consistent with this, peaks in the locus-specific $\omega_{\text{max}}$ align with the focal interaction hubs in the contact map. A complementary graph-based view further shows that H3K27ac-enriched loci cluster in the interaction space and are enriched in high - $\omega_{\text{max}}$ nodes, linking active regulatory hubs to complex viscoelastic signatures. Together, these results indicate that regulatory elements carry distinctive, multi-timescale mechanical fingerprints that can be predicted solely from contact maps.  Heterogeneity observed at larger scales is recapitulated---and refined---at near-nucleosome resolution.

\textbf{Functional implications:} Our findings have important implications for understanding genome function. The multimodal viscoelastic behavior of active chromatin may facilitate the dynamic assembly and disassembly of transcriptional machinery, while the simpler mechanical response of heterochromatin could provide structural stability. The prediction that  relaxation times are encoded in the post-translationally created epigenetic states suggests that chromatin mechanics likely serve as an additional layer of gene regulation, potentially influencing processes such as transcription factor binding, chromatin remodeling, and long-range enhancer-promoter interactions. The mechanical regulation could be particularly important during cellular processes that involve nuclear deformation, such as cell migration, where the viscoelastic properties of different chromatin regions may determine their response to mechanical stress. Additionally, the extension of this approach to biological perturbations, different cell types, and developmental stages could reveal how chromatin mechanics change during cellular differentiation and disease states. For instance, application of our framework to compare wild-type and cohesin-depleted cells could reveal how loop extrusion and cohesin clustering influence chromatin mechanics~\cite{salariLoopExtrusionProvides2025}. Such studies could establish whether chromatin mechanics represents a general principle of genome organization across different biological species.

\begin{figure*}[t]
\centering
\includegraphics[width=.8\linewidth]{figures/schematic.pdf}
\caption{\textsf{Active chromatin regions are associated with enhancer–promoter interactions, transcription factors, coactivators, RNA, and transcriptional condensates, displaying multimode relaxation characteristic of complex viscoelastic responses. In contrast, closed chromatin regions exhibit more compact structures and single-mode relaxation indicative of simpler mechanical dynamics.}}
\label{fig:schematic}
\end{figure*}

The functional implications raised above raise a natural mechanistic question: what are the microscopic interactions that underlie the dynamically varying viscoelastic signatures of active versus closed chromatin? Figure~\ref{fig:schematic} provides a conceptual picture by contrasting compact/closed chromatin with active chromatin in which enhancers and promoters are engaged in transcriptional condensates~\cite{Hnisz17Cell,Sharp22RNA} containing TFs, coactivators, RNA, and associated regulatory elements. Our model is based on \textit{effective} interactions between loci. These interactions may reflect a mix of molecular processes, such as cohesin-mediated loop extrusion or protein bridging. However, because effects are treated implicitly, we cannot separate their specific contributions to chromatin mechanics. In addition, the current framework represents chromatin as a coarse-grained polymer and does not include the surrounding nuclear environment or protein assemblies that can influence local stiffness and dynamics. In particular, transcription factors and coactivator proteins can form condensates \cite{sabariCoactivatorCondensationSuperenhancers2018a, boijaCell2018} that interact with chromatin \cite{shrinivasEnhancerFeaturesThat2019}, potentially altering both its organization \cite{waghCurrentOpinioninStructuralBiology2021} and rheological properties. Nevertheless, we  interpret  the locus-specific viscoelasticity that is encoded in the epigenetic states in chromatin, in light of a scenario where TF/coactivator condensates transiently associate with active chromatin and reshape its effective mechanical environment (Figure~\ref{fig:schematic}). At the JUNB and PRDX2 locus in mESC cells (Figure~\ref{fig:mESC_KLF1_combine}), we find that enhancers, promoters, and gene bodies exhibit increased contact frequency but do not form direct physical contacts or compact clusters in individual structures (Figure~\ref{fig:mESC_KLF1_combine}(d) shows absence of compact clusters formed by active loci). The combination—high contact frequency in population-averaged maps without persistent, compact clustering in single chromatin conformations—is reminiscent of the chromatin--condensate ``wetting'' and/or bridging phenomena described in recent experimental and modeling studies \cite{supakarChromatinBindingRegulates2025, zhaoCondensatedrivenChromatinOrganization2025}. In this picture, transcription factor condensates associate with active chromatin segments and influence their spatial organization and mobility, providing an additional, dynamic source of effective coupling among enhancers, promoters, and gene bodies, and ultimately may regulate gene expression \cite{duDirectObservationCondensate2024}. Through such weak, multivalent interactions, condensates could modulate the local mechanical environment of active chromatin. Although our current framework does not explicitly include these condensates, the distinct viscoelastic signatures of active loci are consistent with the idea that protein condensates and chromatin mechanics are coupled (Figure~\ref{fig:schematic}). Future work integrating explicit condensate--chromatin interactions will be necessary to determine whether the multimodal relaxation behavior predicted here reflects this coupling.

In summary, our study establishes chromatin viscoelasticity as a biophysical property that varies systematically with epigenetic state, providing a new physical perspective on genome organization and function. The findings suggest that mechanical properties may represent an additional dimension of chromatin regulation, with important implications for understanding gene expression and cellular differentiation. The connection between epigenetic state, 3D structure, and mechanical properties reveals a complex interplay that governs genome function, opening new possibilities for understanding and potentially manipulating chromatin behavior in health and disease.

\section{Methods}
\label{sec:methods}

\subsection{Locus-specific viscoelasticity}
\label{method:theory_viscoelasticity}

We developed a theory to predict locus-specific viscoelastic properties of chromatin based on the supposition that knowledge of the static three-dimensional (3D) structure (namely, knowledge of all three-dimensional coordinates, $\{\boldsymbol{r}_i\}$, of the chromatin loci, where $i$ is the locus index) is sufficient to predict the mechanical response of chromatin. The theory is based on the HIPPS-DIMES model that creates a maximum-entropy distribution of the 3D coordinates subject to the constraint that  the Hi-C contact map or imaging data \cite{Shi2021PRX, Shi2023NatComm} be preserved. The maximum-entropy distribution is given by,
\begin{equation}\label{eq:maxent}
P^{\mathrm{MaxEnt}}(\{\boldsymbol{r}_i\})=\frac{1}{Z}\exp\bigg(-\sum_{i<j}k_{ij}||\boldsymbol{r}_i-\boldsymbol{r}_j||^2\bigg),
\end{equation}
\noindent where $Z$ is a normalization constant. The Lagrange multipliers $k_{ij}$ are determined to match the mean pairwise distances between loci $i$ and $j$, $\langle R_{ij} \rangle$, derived from the Hi-C contact probabilities, $P_{ij}$. It was shown previously~\cite{Bintu18Science,Shi2021PRX} that $P_{ij}$  and $\langle R_{ij} \rangle$ are related by a power law.  In Eq. \ref{eq:maxent} the  $k_{ij}$s are the elements in the connectivity matrix, $\boldsymbol{K}$, with $K_{ij}=k_{ij}$ if $i\neq j$ and $K_{ii}=-\sum_{j\neq i}k_{ij}$. By interpreting $k_{ij}$ as effective spring constants in the chromatin network, it is possible to calculate the locus-dependent viscoelastic moduli using standard procedures used in the  Rouse model~\cite{doi1988theory}, see below.

Although the distribution $P^{\mathrm{MaxEnt}}(\boldsymbol{r}_1, \boldsymbol{r}_2, \cdots)$ (Eq.~\ref{eq:maxent}) is calculated using the maximum-entropy principle, we interpret it as a Boltzmann distribution at unit temperature ($k_BT$ is unity, where $k_B$ is the Boltzmann constant and $T$ is temperature) with an effective Hamiltonian, $H=\sum_{i<j}k_{ij}||\boldsymbol{r}_i-\boldsymbol{r}_j||^2$. With this identification, $k_{ij}$ may be thought of as the spring constant between loci $i$ and $j$. Following the theory of dynamics for the Rouse model~\cite{doi1988theory}, the eigenvalue decomposition of the connectivity matrix $\boldsymbol{K}$ is used to calculate the normal modes. Each independent normal mode (indexed by $p$) obeys the Ornstein-Uhlenbeck process.  The relaxation time for mode $p$ is $\tau_p=-\xi/\lambda_p$, where $\xi$ is the friction coefficient and $\lambda_p$ are the eigenvalues of $\boldsymbol{K}$. Within the Rouse model framework~\cite{doi1988theory}, the intrinsic shear relaxation modulus $G(t)$ is expressed as,

\begin{equation}\label{eq:eq1}
G(t) = \frac{c}{N}k_BT \sum_p \exp(-t/\tau_p),
\end{equation}

\noindent where $c$ is the monomer concentration and $N$ is the number of monomers (or loci). Without loss of generality, we set $c=1$ and $k_BT=1$ in our calculation. Applying the Fourier transform yields the frequency-dependent storage modulus $G'(\omega)$ and loss modulus $G''(\omega)$ (where $\omega$ is the angular frequency):

\begin{align}\label{eq:eq2}
G^{\prime}(\omega) = \sum_{p=1} \frac{(\omega\tau_p)^2}{1+(\omega\tau_p)^2}, \
G^{\prime\prime}(\omega) = \sum_{p=1} \frac{\omega\tau_p}{1+(\omega\tau_p)^2}.
\end{align}

The locus-specific storage and loss moduli can be further formulated as,

\begin{align}\label{eq:eq3}
G'_i(\omega) &= \sum_{p=1} v_{pi}^2 \frac{(\omega\tau_p)^2}{1+(\omega\tau_p)^2}, \\
G''_i(\omega) &= \sum_{p=1} v_{pi}^2 \frac{\omega\tau_p}{1+(\omega\tau_p)^2},
\end{align}

\noindent where $v_{pi}$ is the $i$-th component of eigenvector of p-mode $\mathbf{V}_p$ from the eigen-decomposition of $\boldsymbol{K}$ (the contribution of locus $i$ to normal mode $p$). We exclude translation mode ($p=0$) which corresponding to zero eigenvalue $\lambda_0=0$.

These moduli provide a quantitative description of the elastic (solid-like) and viscous (liquid-like) behavior of chromatin across a range of timescales. Although it is known, it is worth stressing that $G'(\omega)$ and $G''(\omega)$ depend on the time scale of motion, thus revealing the scale-dependent relaxation behavior.  To further characterize the frequency-dependent viscoelastic behavior, we also computed the loss tangent, $\tan_{i}(\delta) = G''_i(\omega)/G'_i(\omega)$, that captures the relative contributions of viscous dissipation and elastic storage.

\subsection{Chromatin dynamics subject to mechanical forces}

\label{method:theory_dynamics_force}

The connectivity matrix $\boldsymbol{K}$ that defines the maximum-entropy distribution (Eq.~\ref{eq:maxent}) also governs the overdamped dynamics of the chromatin loci under external forces. In matrix form, the Brownian dynamics equation for the 3D coordinates of $N$ chromatin loci, $\boldsymbol{R}(t) = (\boldsymbol{r}_1(t),\dots,\boldsymbol{r}_N(t))^T$, is

\begin{equation}
\xi \frac{d \boldsymbol{R}}{dt}
= \boldsymbol{K}\,\boldsymbol{R} + \boldsymbol{B} + \boldsymbol{f},
\label{eq:langevin_matrix}
\end{equation}

\noindent where $\xi$ is the friction coefficient, and $\boldsymbol{f}(t)$ is a Gaussian white noise with $\langle \boldsymbol{f}\rangle=\boldsymbol{0}$ and $\langle \boldsymbol{f}(t)\boldsymbol{f}(t')^T\rangle = (2/\beta)\,\xi\,\boldsymbol{I}\,\delta(t-t')$ (we use units with $k_B T=1$, so $\beta=1$). In Eq. \ref{eq:langevin_matrix} \(\boldsymbol{B} = (\boldsymbol{b}_1,\dots,\boldsymbol{b}_N)^{\mathsf{T}}\) is the external forces on each locus, where \(\boldsymbol{b}_i\) is the force vector acting on locus \(i\). For example, to apply a constant force \(F\) along the positive \(x\)-direction to a single locus \(i\) while leaving all other loci unforced, one sets \(\boldsymbol{b}_i = (F,0,0)\) and \(\boldsymbol{b}_j = \boldsymbol{0}\) for all \(j \neq i\).

We diagonalize $\boldsymbol{K}$ as $\boldsymbol{K} = \boldsymbol{V}\,\boldsymbol{\Lambda}\,\boldsymbol{V}^T$, where $\boldsymbol{\Lambda}=\mathrm{diag}(\lambda_1,\dots,\lambda_N)$ contains the eigenvalues $\lambda_p\le 0$ and the columns of $\boldsymbol{V}$ are the corresponding orthonormal eigenvectors. These are the same eigenmodes used to compute the relaxation times $\tau_p=-\xi/\lambda_p$ and the viscoelastic moduli in Eqs.~\eqref{eq:eq2}–\eqref{eq:eq3}. We define normal-mode coordinates and transformed forces/noise as $\boldsymbol{X}=\boldsymbol{V}^T\boldsymbol{R}$, $\boldsymbol{B}'=\boldsymbol{V}^T\boldsymbol{B}$, and $\tilde{\boldsymbol{f}}=\boldsymbol{V}^T\boldsymbol{f}$. Substituting into Eq.~\eqref{eq:langevin_matrix} gives

\begin{equation}
\xi \frac{d \boldsymbol{X}}{dt}
= \boldsymbol{\Lambda}\,\boldsymbol{X} + \boldsymbol{B}' + \tilde{\boldsymbol{f}},
\label{eq:langevin_modes_vector}
\end{equation}

\noindent and the orthogonality of $\boldsymbol{V}$ implies that $\tilde{\boldsymbol{f}}$ has the same statistics as $\boldsymbol{f}$, so the modes are dynamically independent.
Each component $X_p$ obeys a scalar Ornstein–Uhlenbeck (OU) process with a constant offset~\cite{Hyeon17JPCB},

\begin{equation}
\frac{d X_p}{dt}
= -\theta_p X_p + \theta_p \mu_p + \sigma\,\eta_p(t),
\label{eq:OU_standard}
\end{equation}

\noindent where $\eta_p(t)$ is white noise. Comparing Eq.~\eqref{eq:OU_standard} with Eq.~\eqref{eq:langevin_modes_vector} yields
$\theta_p = -\lambda_p/\xi$, $\sigma = \sqrt{2/(\xi\beta)}$, and $\mu_p = -B'_p/\lambda_p$ for all nonzero modes ($\lambda_p\neq 0$). In matrix notation, this can be written compactly as $\boldsymbol{\theta} = -\boldsymbol{\Lambda}/\xi$, $\sigma = \sqrt{2/(\xi\beta)}$, and $\boldsymbol{\mu} = -\boldsymbol{\Lambda}^{-1}\boldsymbol{V}^T\boldsymbol{B}$, with the zero eigenvalue associated with global translation removed (or the center-of-mass fixed). Thus, the external force shifts the mean of each nontrivial mode from $0$ to $\mu_p$, while the relaxation times $\tau_p=-\xi/\lambda_p$ are fully determined by the spectrum (eigendecomposition) of $\boldsymbol{K}$.

For numerical calculations of chromatin dynamics subject to external forces, we work entirely in mode space. We compute the eigendecomposition $\boldsymbol{K}=\boldsymbol{V}\boldsymbol{\Lambda}\boldsymbol{V}^T$ once for a given connectivity matrix $\boldsymbol{K}$, then, at each time step, project the external force profile onto the modes ($\boldsymbol{B}'=\boldsymbol{V}^T\boldsymbol{B}$), update each independent OU mode according to either the Euler–Maruyama integration scheme or the exact OU propagator \cite{gillespiePhys.Rev.E1996}, and finally transform back to real space via $\boldsymbol{R}=\boldsymbol{V}\boldsymbol{X}$. This procedure provides an efficient and numerically stable method to simulate chromatin dynamics with external forces and to relate these perturbations to the locus-specific viscoelastic properties.

\subsection{Fractal dimension $D_f$}
\begin{figure}[t]
\centering
\includegraphics[width=.8\linewidth]{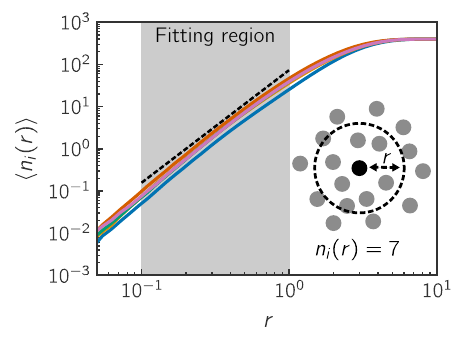}
\caption{\textsf{Ensemble-averaged neighbor count for locus $i$, $\langle n_i(r)\rangle$, as a function of radius $r$.
The scaling regime $0.1 \le r \le 1$ is used to fit $\langle n_i(r)\rangle \propto r^{D_f}$ to extract the locus-specific fractal dimension $D_f$.
Curves for several representative loci are shown in different colors.
The dashed line indicates the expected power-law scaling (schematic).}}
\label{fig:gm12878_neighbor_counts_vs_distance_schematic}
\end{figure}

We quantified the locus-specific fractal dimension $D_f(i)$ from the scaling of the number of neighbors within a spatial radius $r$. For each genomic region, we generated an ensemble of $M=10{,}000$ independent 3D structure realizations. For each structure and each locus $i$, we computed the number of loci within a sphere of radius $r$ centered at locus $i$,
\begin{equation}
n_i(r)=\sum_{j\neq i} H\!\left(r-r_{ij}\right),
\end{equation}
\noindent where $r_{ij}$ is the distance between locus $i$ and $j$, $H(x)$ is the Heaviside step function, $H(x)=1$ if $x\ge 0$ and $0$ otherwise. We then computed the ensemble-averaged local number of neighbors profile for each locus, $\langle n_i(r)\rangle=(1/M)\sum_{a=1}^{M} n_i^{(a)}(r)$, where $a$ specifies the structures in the ensemble. Assuming fractal-like packing over an intermediate length-scale window, we determined $D_f(i)$ by fitting $\langle n_i(r)\rangle$ to the power-law form, $\langle n_i(r)\rangle \propto r^{D_f(i)}$ over the range $0.1 \le r \le 1$ (in simulation length units), see Figure~\ref{fig:gm12878_neighbor_counts_vs_distance_schematic}.

\subsection{Local stiffness, $\kappa_i$}

We calculate the local mechanical stiffness ($\kappa_i$) of each locus based on the sensitivity to a local perturbation, assessed via the changes to the mode spectrum and the positional fluctuation \cite{zhengStructure2005, vuSci.Adv.2022}. We impose a small perturbation that rescales all the  off-diagonal coupling $k_{ij}$ associated with the locus $i$ by a factor $(1+\alpha)$. The first-order change in the eigenvalue $\lambda_p$
with respect to $\alpha$ is,

\begin{equation}
\delta\lambda_p(i)
\equiv
\left.\frac{\mathrm{d}\lambda_p}{\mathrm{d}\alpha}\right|_{\alpha=0}
= -\sum_{j \neq i }^N k_{ji}\,\bigl(v_{pj} - v_{pi}\bigr)^2,
\end{equation}

\noindent where $v_{pi}$ is the $i$-th component of eigenvector of p-mode $\mathbf{V}_p$. $\delta\lambda_p(i)$ measures the sensitivity of mode $p$ to local mechanical changes at locus
$i$.

To obtain mechanical sensitivity for each locus, we average over modes with inverse-eigenvalue weights, $\delta\lambda_i = \sum_{p>0} \delta\lambda_p(i) / \lambda_p$. The mean squared positional fluctuation of locus \(i\) is $\langle R_i^2 \rangle = -3 \sum_{p>0} \lvert v_{pi}\rvert^2/ \lambda_p$. The local stiffness (or mechanical importance) of locus $i$ is defined
using,

\begin{equation}
\kappa_i = \frac{\delta\lambda_i}{\langle R_i^2 \rangle},
\end{equation}

\noindent which is a measure of how strongly a local mechanical perturbation at locus $i$ modifies the global relaxation spectrum, normalized by the amplitude of its thermal fluctuations.

\subsection{Hi-C/Micro-C Data Processing}
\label{method:hi_c_processing}
We used the Hi-C data from human GM12878 lymphoblastoid cells and Micro-C for mouse embryonic stem cells (mESCs). For GM12878, we analyzed five genomic regions: Region 1: Chromosome 1, 31-41 Mb; Region 2: Chromosome 6, 28-38 Mb; Region 3: Chromosome 7, 27-37 Mb; Region 4: Chromosome 11, 32-42 Mb; Region 5: Chromosome 17, 47-57 Mb. Each region was analyzed at 25 kb resolution. Hi-C contact maps were processed using the HIPPS-DIMES framework~\cite{Shi2021PRX,Shi2023NatComm} to generate the connectivity matrices representing the inter-loci interaction network. For mESC, we analyzed Chromosome 8: 85.68 Mb to 85.82 Mb region at near nucleosome-level resolution (250 bp).

All experimental contact maps (Hi-C and Micro-C) were first balanced (Knight-Ruiz matrix balancing) and then processed using the neighbor balancing method \cite{Paggi2025}. This method  corrects for the uniform spatial density assumed by the standard Hi-C balancing method.

\section*{Acknowledgments}
We are grateful to Bhuwan Poudel for useful comments and discussions. This work was supported by a grant from the National Science Foundation (CHE 2320256) and the Welch Foundation through the Collie-Welch Chair (F-0019). 

\section*{Author contributions}
G.S. and D.T. designed research; G.S. and D.T. performed research; G.S. and D.T. analyzed data; G.S. and D.T. wrote the paper.

\section*{Data availability}
The Hi-C data for human GM12878 cells~\cite{harrisNatCommun2023} were obtained from ENCODE (accession: ENCFF065LSP), and RCMC data for mouse embryonic stem cells (mESCs)~\cite{GoelNatGenet2023} were obtained from GEO (accession: GSE207225).
We used the raw BAM files for ChIP-seq data for GM12878 from ENCODE (H3K27ac: ENCSR000AKC, H3K27me3: ENCSR000AKD, H3K36me3: ENCSR000AKE, H3K9me3: ENCSR000AOX, H3K4me3: ENCSR000AKA, H3K4me1: ENCSR000AKF). 
Read counts were computed from replicate BAM files using \texttt{BEDTools} commands \texttt{makewindows} and \texttt{multicov} to generate fixed-size genomic bins, and normalized to reads per million (RPM) based on total mapped reads estimated with \texttt{SAMtools}. 
The resulting files provide combined replicate counts and RPM-normalized ChIP-seq signal at user-specified resolutions.
For mESCs, we processed the H3K27ac ChIP-seq data (GSE90893) following the procedure described in Goel \textit{et al.}~\cite{GoelNatGenet2023}. The original mm9-aligned signal was mapped to the mm39 genome using CrossMap~\cite{zhaoCrossMapVersatileTool2014} and subsequently averaged into 250~bp bins.

\section*{Code availability}
The code for HIPPS-DIMES model presented in this work and the detailed user instructions can be accessed at the GitHub repository https://github.com/anyuzx/HIPPS-DIMES. The data analysis is performed using Python 3.12 in Jupyter Lab. The Python packages used in data analysis are Scipy, Numpy, and Pandas. Visualizations of chromatin structures are rendered using VMD \cite{humphrey1996vmd}.

\newpage 
\section{Appendix A: HIPPS-DIMES reconstructed contact maps for GM12878}
\begin{figure*}[t]
\centering
\includegraphics[width=\linewidth]{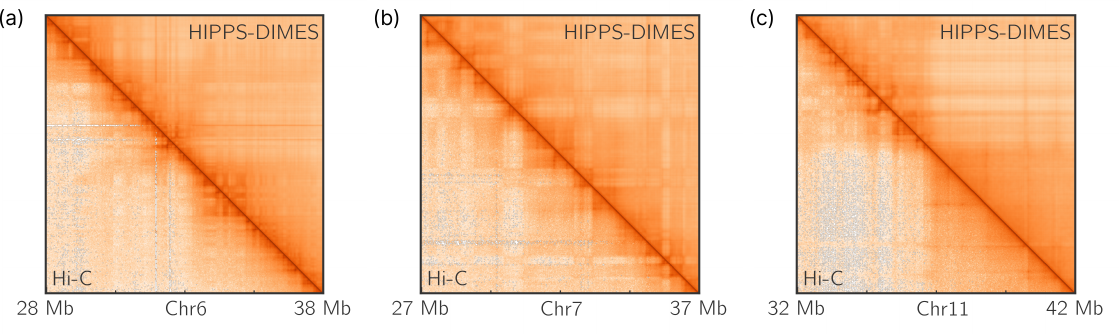}
\caption{\textsf{Comparison between Hi-C and HIPPS-DIMES contact maps in GM12878. (a) Chr6: 28--38 Mb. (b) Chr7: 27--37 Mb. (c) Chr11: 32--42 Mb.}}
\label{fig:gm12878_10mb_contact_maps_combine}
\end{figure*}

Figure~\ref{fig:gm12878_10mb_contact_maps_combine} compares Hi-C and HIPPS-DIMES contact maps for three 10 Mb regions in GM12878: Chr6 (28--38 Mb), Chr7 (27--37 Mb), and Chr11 (32--42 Mb).

\section{Appendix B: Viscoelastic properties of Chromosome 3 in GM12878}
\begin{figure*}[t]
\centering
\includegraphics[width=.9\linewidth]{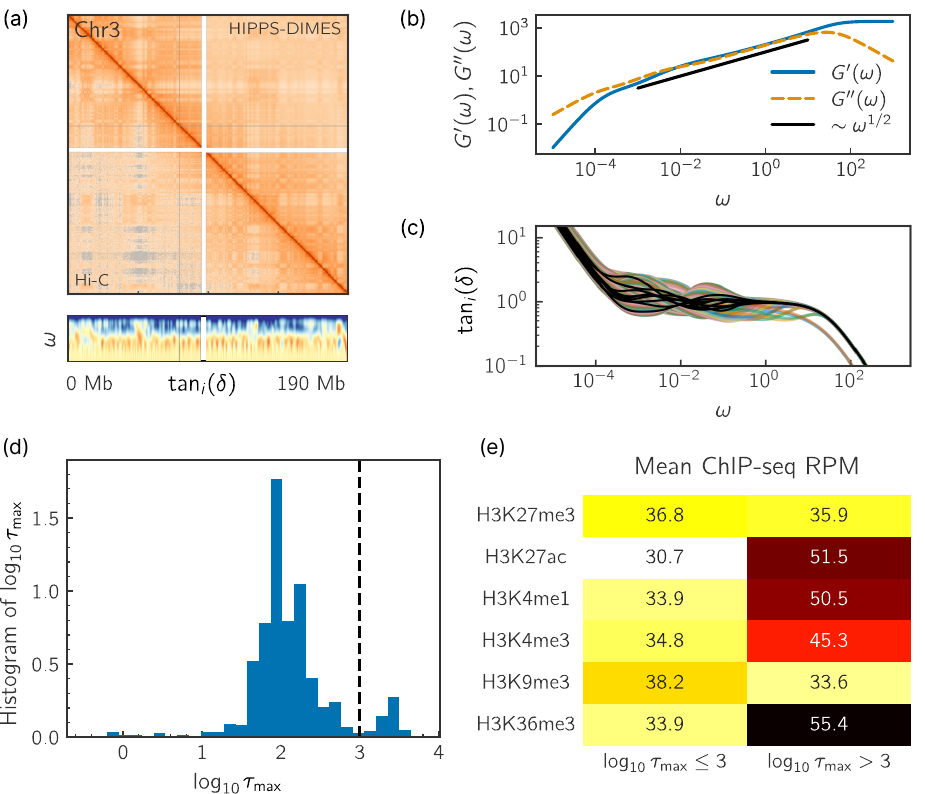}
\caption{\textsf{Viscoelastic properties of Chromosome 3 (Chr 3) in GM12878 and the association with epigenetics. (a) Comparison of contact map of Chromosome 3 between Hi-C and HIPPS-DIMES. Bottom panel shows the $\tan_i(\delta)$ heatmap. (b) Storage and loss modulus for Chr 3 as a whole. Solid lines are $G^{\prime}(\omega)$ and dashed lines correspond to $G^{\prime\prime}(\omega)$. (c) $\tan_i(\delta) = G_{i}^{\prime\prime}(\omega)/G_{i}^{\prime}(\omega)$. Black solid lines are for a few randomly selected loci. (d) Histogram of the largest relaxation time $\tau_{\text{max}}$. Vertical dashed line marks the  separation of two subpopulations. (e) Average histone modification levels across the two groups. Loci with long relaxation times are enriched in active chromatin marks (H3K27ac, H3K4me1, H3K4me3, H3K36me3).}}
\label{fig:gm12878_chr3_combine}
\end{figure*}

Figure~\ref{fig:gm12878_chr3_combine} illustrates the application of our framework to the whole Chromosome 3 in GM12878. We applied HIPPS-DIMES to the whole Chromosome 3 Hi-C contact map at 100 kb resolution. We computed both chromosome-averaged and locus-specific viscoelastic moduli, $G'(\omega)$ and $G''(\omega)$, and the loss tangent $\tan_i\delta(\omega) = G''_i(\omega)/G'_i(\omega)$, revealing pronounced heterogeneity in the viscoelastic behavior across loci and frequencies. From the mode spectrum, we defined for each locus an effective largest relaxation time $\tau_{\mathrm{max}}(i)$ and examined the associated distribution along the chromosome, which enables us to classify the loci into slow and fast relaxation groups. Comparing histone modification profiles between these groups shows that loci with long relaxation times are enriched in active marks (H3K27ac, H3K4me1, H3K4me3, H3K36me3).

\clearpage

\newpage
\bibliography{references}

\end{document}